\documentclass{aa}

\usepackage{psfig}

\begin{document}
\thesaurus { }

\title{ESO~603-G21: A strange polar-ring galaxy\thanks{Based 
on observations made at the  Observat\'{o}rio  do Pico  dos  Dias  (OPD),  
operated  by the MCT/Laborat\'{o}rio Nacional de Astrof\'{\i}sica, 
Brazil}}

\author{V.P.  Reshetnikov\inst{1,3}\and  M.  Fa\'{u}ndez-Abans\inst{2}
\and     M.    de     Oliveira-Abans\inst{2}\\     resh@astro.spbu.ru,
mfaundez@lna.br, mabans@lna.br }

\offprints{V.P. Reshetnikov~resh@astro.spbu.ru}   

\institute{Astronomical Institute of St.Petersburg State
University, 198504 St.Petersburg, Russia
\and MCT/Laborat\'{o}rio   Nacional  de   Astrof\'{\i}sica,   
Caixa  Postal  21, CEP:37.504-364, Itajub\'{a}, MG, Brazil
\and Isaac Newton Institute of Chile, St.Petersburg Branch}  

\date{Received 16 August 2001/Accepted 26 November 2001}

\titlerunning{Photometry of ESO~603-G21}

\maketitle

\markboth{Photometry of ESO~603-G21}{Reshetnikov, Fa\'{u}ndez-Abans, 
de Oliveira-Abans}

\abstract{ 
We present the results of $B$, $V$, $R$ surface  photometry of  ESO~603-G21
-- a galaxy with a possible polar ring.  The morphological  and photometric
features of this galaxy are  discussed.  The  central  round  object of the
galaxy is rather red and presents a nearly
exponential  surface  brightness  distribution.  This central  structure is
surrounded  by a blue warped ring or disk. The  totality  of the  observed
characteristics  (optical and NIR colors,  strong color  gradients,  HI and
H$_2$  content, FIR  luminosity  and  star-formation  rate,  rotation-curve
shape, global mass-to-luminosity ratio, the agreement with the Tully-Fisher
relation,  etc.)  shows that  ESO~603-G21  is  similar  to  late-type
spiral galaxies.  We suppose that morphological  peculiarities and the possible
existence  of  two   large-scale   kinematically-decoupled   subsystems  in
ESO~603-G21  can be explained as being a result of dissipative  merging of
two spiral  galaxies or as a consequence  of a companion  accretion  onto a
pre-existing spiral host. 
}

%\maketitle

\authorrunning{Reshetnikov, Fa\'{u}ndez-Abans, de Oliveira-Abans}

\section{Introduction}

The past several years have been very rich in observational studies of
galaxy   formation  and   evolution.  One  of  the  most   interesting
conclusions  of these studies is the  continuous  assembly of galaxies
(see Ellis 2001 for a recent  review).  Among the best local  examples
of delayed galaxy formation are the so-called  multi-spin  galaxies --
objects with more than one axis of rotation  (Rubin 1994).  Polar-ring
galaxies  (PRG) are  probably  the best known  instance of  multi-spin
objects (see Whitmore et al.  1990,  hereafter PRC, for definition and
catalog of such objects ).  PRG probably represent  products of merger
or external accretion phenomena (PRC, Reshetnikov \& Sotnikova 1997).

In this article we present the results of photometric  observations of
ESO~603-G21  -- a good PRG  candidate  according  to  Whitmore  et al.
(1990) (see Fig.~3f in Whitmore et al.  1990 and contour maps  in
our  Fig.~2). This galaxy  resembles  an  early-type
galaxy with a well-developed  bulge and an extended warped and edge-on
disk/ring.  A dust lane can be seen at the  intersection  of the bulge
and the disk/ring.

The  spectroscopic  data for this object indicate a complex  scenario.
The rotation curves for \mbox{ESO~603-G21} show that the gas and stars
in the disk/ring  revolve around the minor axis (PRC, Arnaboldi et al.
1994).  At  P.A.~=~24$^{\rm  o}$ (minor  axis),  the  spectra  show no
motion of the gas  perpendicularly to the disk/ring.  In contrast, the
absorption  line rotation  curve  indicates  the  existence of stellar
motion  along  this  axis   (Arnaboldi  et  al.  1994).  There  stellar
kinematics  possibly  indicate  that the  underlying  stellar body is
triaxial (Arnaboldi et al.  \cite{a1}, Arnaboldi et al.  \cite{a}).

\section{Observations and reductions}

The observations were performed with the 1.6-m telescope at the OPD on
August 2, 2000,  equipped with direct  imaging  camera \#1 (details in
{\tt  http:  //www.lna.br/instrum/camara/camara.html})  and CCD  \#106
(1024x1024 square pixels, 24 $\mu$m each), with  RON~=~4.1$e^{-}$  and
gain~=~5.0$e^{-}$/ADU.

The data were acquired with standard Johnson $B$, $V$ and Cousins $R$,
$I$  filters.  The  details  of  the  observations  are given in  Table  1.
Photometric  calibration  was  accomplished  using standard stars from
Landolt (1992).  We have used the mean extinction coefficients for the
OPD:  0.$^m$34,  0.$^m$19  and  0.$^m$14  in  the  $B$,  $V$  and  $R$  
passbands,
respectively .  The $I$-band image was obtained under  non-photometric
conditions  and so we have used it to  determine  the  position of the
galaxy  nucleus  only.  Reduction of the raw CCD data has been made in
the standard  manner using the  ESO-MIDAS\footnote{MIDAS  is developed
and maintained by the European Southern Observatory.}  package.

\begin{table}
\caption{Observation log for August 2, 2000}
\begin{center}
\begin{tabular}{cccccc}
\hline
Bandpass & Airmass  & Exp. & Seeing & Sky \\
         &          &(sec) &  ($''$)& mag. \\
\hline
$R$      & 1.147    & 900 &   1.0   &  20.4 \\
$R$      & 1.109    & 900 &   1.0   &  20.4  \\
$R$      & 1.078    & 900 &   1.0   &  20.4  \\
       &  &  &  &  \\
$B$      & 1.052    & 900 &   1.8   &  21.5 \\
$B$      & 1.033    & 900 &   1.7   &  21.5   \\
$B$      & 1.018    & 900 &   1.8   &  21.5  \\
&  &  &  &  \\
$V$      & 1.008    & 900 &   1.7   &  20.7 \\
$V$      & 1.002    & 900 &   1.7   &  20.7  \\
&  &  &  &  \\
$I$      & 1.001    & 900 &   1.7   & Clouds \\
\hline
\end{tabular}
\end{center}
\end{table}

\section{Results}

\subsection{Integrated photometry}

The total  magnitude  ($B_{\rm  T}=15.0 \pm 0.1$) found by us from the
multiaperture   photometry   is   somewhat   brighter   than  that  of
NED\footnote{NASA/IPAC  Extragalactic  Database.}  (15.3)  and  of PRC
(15.5).  Our $R$ magnitude  14.1\,$\pm$\,0.1  is in agreement with the
PRC value $m_R=14.07$.

Some results of color  measurements  are presented in Arnaboldi et al.
(1994).  According  to those  authors,  the $B-R$  color  index in the
galaxy  center is +2, while the outer  regions are rather  bluer, with
$B-R \approx 1$.  Our values (see Figs.~3 and 4) are in good agreement 
with those results.

\subsection{Optical morphology}

In order to enhance possible internal  structures of this galaxy, some
tests have been carried out  following  the  experiments  performed by
Fa\'{u}ndez-Abans and de Oliveira-Abans (\cite{fa}) and employing some
IRAF tasks.  For illustrative  purposes, Fig.~1a displays the residual
image of the subtraction  of a 60 x
60-pixel median-filtered kernel from the original $R$ frame.  An inner
elongated bright stellar  component,  around another yet smaller round
off-center  component have been  enhanced.  The dark dust lane (ring?)
probably  immersed in the warped  material  external to the disk/ring,
and some faint clumps to the east have also been enhanced.

\rm 

%----------------------> figure 1
\begin{figure}
\vbox{
\includegraphics{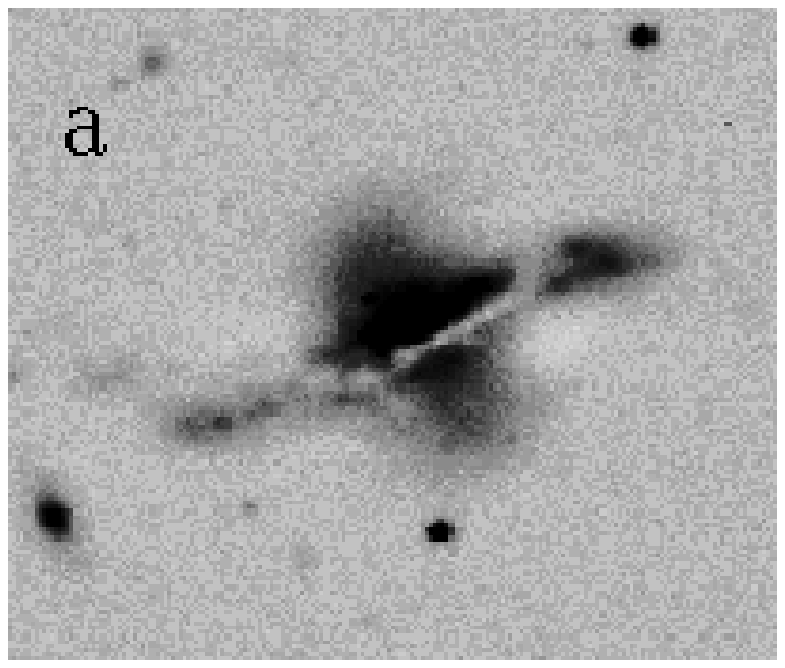}
\includegraphics{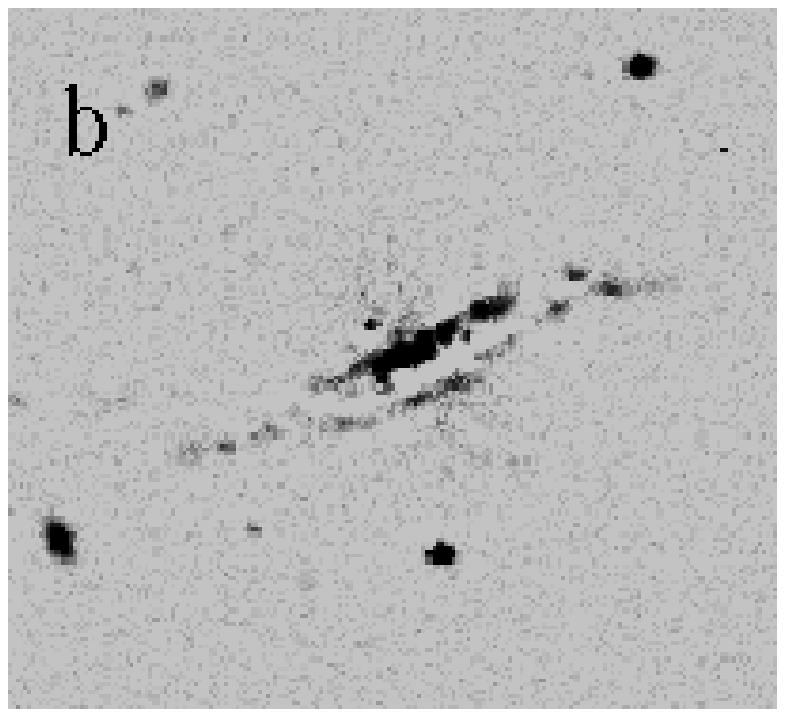}
\includegraphics{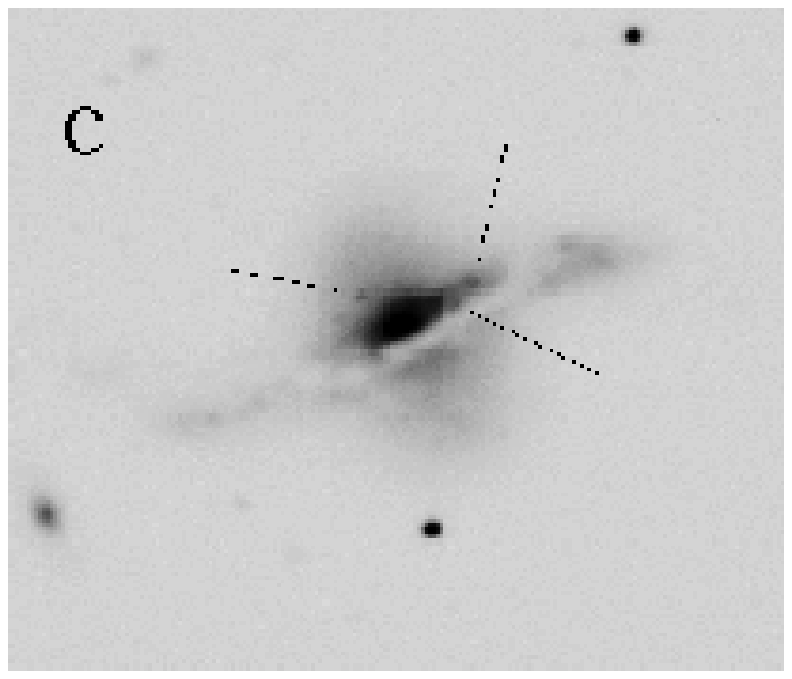}
}%\par
\vspace*{18.0cm}
\caption[1]{{\bf (a)} -- Inverse gray scale residual image of the subtraction
of the 60 x 60-pixel median-filtered $R$ frame from the original $R$ image.
North is to the top and east is on the left. The image size is 
224\arcsec x 200\arcsec. {\bf (b)} -- The same as {\bf (a)} for a 10~x~10-pixel 
kernel. {\bf (c)} -- Enhancement by transform processing and high-pass 
filtering of the $R$ image.  The  mark-lines  are on purpose  faint and 
poor so as not to cover the image.}
\end{figure}

The use of a median 30~x~30-pixel kernel filter has enhanced the fuzzy
material  aligned  with the  apparent  major  axis of the  bulge.
Interestingly  enough,  after  the  application  of  a  10 x
10-pixel  kernel median filter, this fuzzy  material turned out to
be filaments (see Fig.~1b).  This result is in agreement with Arnaboldi
et al.  (\cite{a}).  The ``five to ten" filaments so-found seem to lie
perpendicular to an almost edge-on  exponential  disk.  This is indeed an
interesting PR dusty galaxy.

Other  features  have also been  revealed  through  a  high-pass  filtering
transform  processing of the original $R$ frame:  clumps  (satellites?)  in
the  bulge  and  within  the  polar  ring, a faint  and  underlying  smooth
component (in the central part of the bulge), and an  off-center  component
inside  the  central   structure.  The  most  prominent  clumps  have  been
indicated in Fig.~1c by thin lines.  Careful visual  inspection of the image
display (SAO's ds9) shows that the filaments tails point towards the West,
superimposed on faint smooth regions that we call lobes.

\subsection{Global photometric structure}

Contour  maps  of  ESO~603-G21  are  presented  in  Fig.~2.  The  main
features  of  the  galaxy  are  clearly  visible:  a  main  body  with
approximately  round isophotes and an almost edge-on warped  structure
(polar ring?)  crossing the central object and strongly distorting the
surface brightness  distribution.  The galaxy is surrounded by a faint
halo whose major  axis is  aligned  with  the  major  axis of the
possible ring.

%----------------------------Figure 2
\begin{figure*}[!ht]
\psfig{file=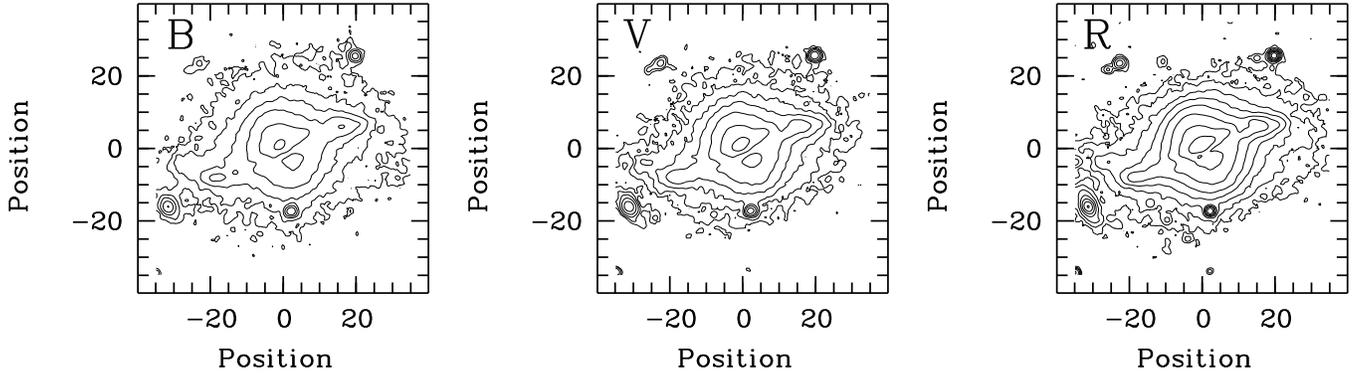,width=18.0cm,angle=-90,clip=}
\caption[2]{Contour  maps  of  ESO~603-G21  in the  $B$,  $V$  and  $R$
passbands.  The  faintest  contours in the $B$, $V$, and $R$ bands are
25.2, 25.3, and 25.4  mag~arcsec$^{-2}$,  respectively.  The isophotes
are separated by 0.75 mag.  The large tickmarks are 20$\arcsec$ apart.
North is to the top and east is on the left.}
\end{figure*}

Fig.~3a,c display the surface brightness profiles of ESO~603-G21 along
the major and minor  axes.  In the major  axis  profile  the  possible
disk/ring  is seen as two  symmetrical  ``bumps''  at $r  \approx  \pm
20''$.  At the SW  part  of  the  profile  along  the  minor  axis  at
$r\approx  -2''$, a  depression  due to  absorption  in the  ring/disk
projected here onto the central part of the galaxy is seen.

%-------------------------- figure 3
\begin{figure*}[!ht]
\centerline{\psfig{file=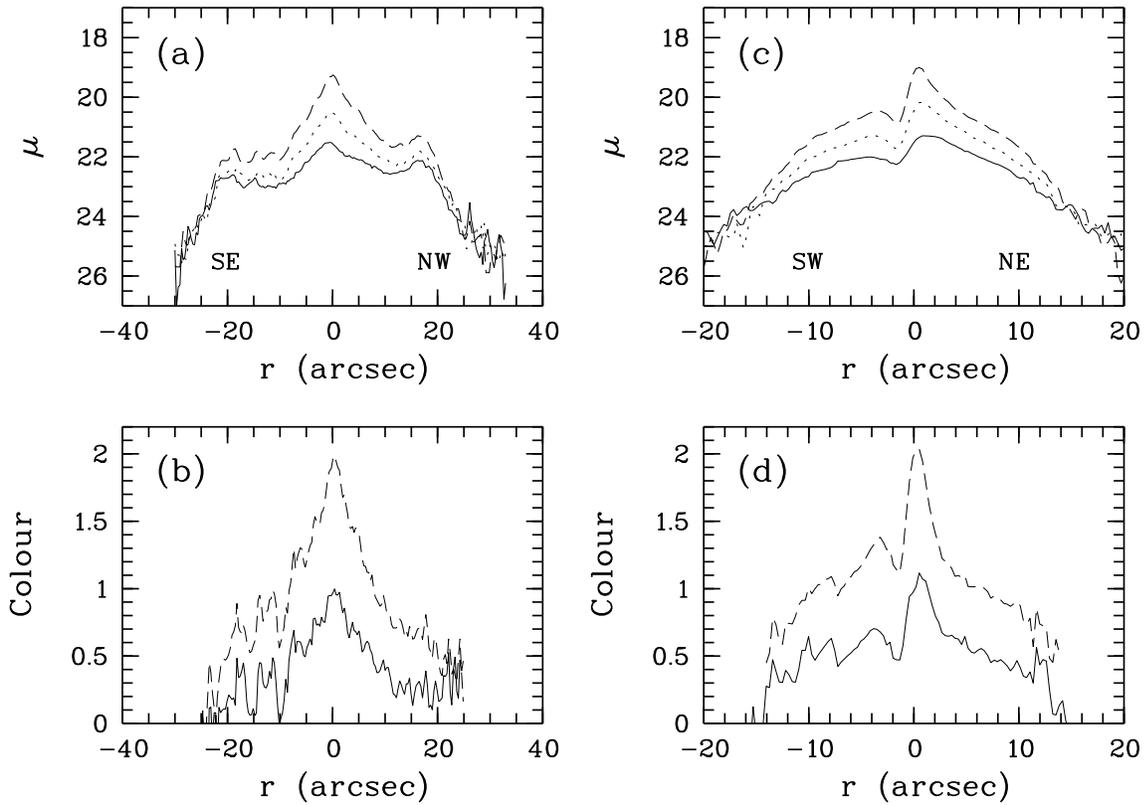,width=16.0cm,angle=-90,clip=}}
\caption[3]{Photometric  profiles  for  ESO~603-G21:  (a), (b) -- along
the major axis (P.A.=114$^{\rm  o}$); (c), (d) -- along the minor axis
(P.A.=24$^{\rm  o}$).  Solid  lines  in  (a)  and  (c)  represent  the
distributions  in the $B$  passband,  dotted  lines in $V$, and dashed
ones in the $R$.  Solid lines in (b) and (d) show the  distribution of
the $B-V$ color, dashed in the $B-R$ color.}
\end{figure*}

Excluding the regions of the ``bumps''  ($\mid  r\mid\,=\,10\arcsec  -
25\arcsec$), the surface brightness  distribution along the major axis
may be approximated by that of an exponential disk (see also Arnaboldi
et al.  \cite{a}).  In the $R$ passband the disk characteristics  are:
$\mu_0^0(R)=19.55$  (corrected for galactic  absorption)  and $h=1.22$
kpc.  The disk of the galaxy is thus  relatively  bright and  compact.
The minor axis  profile in the $R$ filter is also  approximated  by an
exponential one with $h=3\farcs  5$\,=\,0.73  kpc (Fig.~3).  The total
magnitude of the central  exponential object is $R=14.9$  (assuming an
apparent axis ratio  $b/a=1.0$).  Therefore, the ratio of luminosities
of the central round object  (bulge?)  to the  disk/ring is $\sim$1 in
the $R$ passband.

Fig.~3b,d display the behavior of the observed color indices along the
major and minor axes.  Both profiles show very strong color gradients:
the central parts of the galaxy are red ($B-V  \approx  +0.8-1.0,  B-R
\approx  +1.5-2.0$),  while  the  outer  ones are blue  ($B-V  \approx
+0.2-0.5, B-R \approx  +0.5-1.0$).  The galaxy disk/ring is very blue:
$B-V \approx  +0.2-0.3$ and $B-R \approx +0.5$ at $r \approx \pm 20''$
along the  major  axis  (where  the two  ``bumps''  are  visible).  In
Fig.~3d, the region of the disk/ring  projection exibits a local color
minimum,  thus  supporting  our  conclusion  about the blueness of the
ring.

Fig.~4 gives the 3D  distribution  of the observed color index $B-R$ within
the central  region of the galaxy.  In this  figure, the  disk/ring  is the
notably blue path (narrow  ``valley")  crossing  the
central region. A prominent color gradient is evident in the figure.

In order to study the galaxy structure in the near-infrared  (NIR) spectral
region, we have  extracted the $J$, $H$ and $K$ images of  {\mbox
ESO~603-G21}  from the second  incremental  data release of the Two Micron
All    Sky     Survey     (Skrutskie     et    al.    1997;     see    {\tt
http://www.ipac.caltech.edu/2mass}).  The NIR colors of the  galaxy  (Table
2) is
usual for spiral  galaxies  (see  Fig.~9 in Iodice et al.  2001).  We found
that to a first  approximation  the galaxy  structure can be described as a
thick double  exponential  disk with strong color gradients along the major
and minor axes.  In Table 3 we present the scalelength  ratios in different
color  bands,  both  along the major and minor  axes.  The  large  observed
ratios are typical  for dusty  late-type  spiral  galaxies  (e.g.  de Grijs
1998).

\begin{table}
\caption{General properties of ESO~603-G21}
\begin{tabular}{lll}
\hline
Parameter &  Value & Ref. \\
\hline
Morphological type             & Sbc          &  NED   \\
Heliocentric systemic velocity & 3\,150 km/s    &  PRC   \\
Distance                       & 42.9 Mpc     &      \\
                               & (1$''$=208 pc)&     \\
Redshift                       & 0.01042      &  NED  \\
P.A.                           & 114$^{\rm o}$ &                 \\
Major axis, D$_{25}$~($\mu_B=25$) & 55$''$ (11.4 kpc) &               \\
Axial ratio, $b/a~(\mu_B=25)$   & 0.7           &              \\
Inclination, $i$               & 80$^{\rm o}$:  &            \\
                               &                &             \\
Total apparent                 &                          &      \\
magnitudes and colors:         &                          &      \\
$B_{\rm T}$                          & 15.0$\pm$0.1 &      \\
$(B-V)_{\rm T}$                      & +0.30$\pm$0.05 &   \\
$(V-R)_{\rm T}$                      & +0.60$\pm$0.05 &   \\
$(J-H)_{\rm 2MASS}$            & +0.86          & [1]       \\
$(H-K)_{\rm 2MASS}$            & +0.33          & [1]       \\
Galactic absorption ($B$-band) & 0.14           & [2] \\
Internal absorption ($B$-band) & 1.0            &     \\
Absolute magnitude, $M_B^0(0)$ & --19.3         &     \\
                               &                &     \\
Exponential disk:              &                &     \\
major axis:                    &                &     \\
$\mu_0^0(R)$                   & 19.55          &     \\
$h$               & 5.$''$9 (1.22 kpc)  &     \\
minor axis:                    &        &     \\
$h$                            &  3.\arcsec5 (0.73 kpc)  &   \\
Exponential central object:    &                &      \\
$R_{\rm T}$                    &  14.9          &      \\
$L_R^{\rm central ~object}$ / $L_R^{\rm disk/ring}$  &  $\sim$~1   &  \\ 
                               &                     &        \\
M(HI)& 6.2\,10$^9$ M$\odot$ & [3,4] \\
M(H$_2$)& 1.1\,10$^9$ M$\odot$ & [5] \\
M(HI)/$L_B^0(0)$               & 0.76 M$\odot$/$L_{\odot,B}$   &   \\
M(H$_2$)/M(HI)                 & 0.18                 &      \\
HI linewidth, $W_{20}$         & 286 km/s  & [4] \\
HI linewidth, $W_{50}$         & 251 km/s  & [4] \\
                               &                      &     \\
Far-infrared luminosity, $L_{\rm FIR}$ & 4.85\,10$^9$ $L_{\odot}$ & NED, [3] \\
Far-infrared color, f$_{60}$/f$_{100}$ & 0.50   & NED \\
Mass of dust, M$_d$            & 1.5\,10$^6$ M$\odot$ & [6]  \\
SFR$_{\rm FIR}$                & 2.5 M$\odot$/yr  & [7]  \\
SFE (=$L_{\rm FIR}$/M(H$_2$))  & 4.4 $L_{\odot}$/M$_{\odot}$ \\
\hline 
\end{tabular} 
[1] -- Skrutskie et al. 1997;
[2] -- Schlegel et al. 1998; [3] -- Richter et al. 1994; 
[4] -- van Driel et al. 2000; [5] -- Galletta et al. 1997;
[6] -- Young et al. 1989; [7] -- Hunter et al. 1986
\end{table}

\begin{table}
\caption{Exponential  scalelength ratios}
\begin{center}
\begin{tabular}{|c|c|c|}
\hline
Gradient & Major axis  & Minor axis \\
\hline
$h_R/h_K$& 1.9$\pm$0.3 & 1.8$\pm$0.2 \\
$h_J/h_K$& 1.3$\pm$0.2 & 1.7$\pm$0.2 \\
$h_H/h_K$& 1.4$\pm$0.3 & 1.4$\pm$0.3 \\
\hline
\end{tabular}
\end{center}
\end{table}

Table 2 summarizes the main characteristics of ESO~603-G21, both found
in this  work and  collected  from the  literature.  The  last  column
provides  the  corresponding   references,  where  the  absence  of  a
reference  indicates that the given value has been  determined in this
work.

\subsection{Dust and internal absorption}

The mass of warm ($T_d \approx  35$\,K) dust found from the 100 $\mu$m
IRAS flux is  1.5\,10$^6$  M$\odot$  (Table  2).  Assuming  that  only
$\sim$~10\%\,--\,20\% of the dust mass in disk galaxies is warm enough
to radiate in the IRAS bands  (Devereux  \& Young  \cite{dy}),  we can
estimate the total dust mass in the galaxy as $\sim$10$^7$ M$\odot$.

What is the total internal absorption in ESO~603-G21? The standard empirical
description of the extinction as a function of galactic inclination is 
\begin{center}
$A(i)\,=\,C$\,log$(a/b)$, 
\end{center}
where the extinction parameter $C$ depends
on the morphological type, total luminosity, and on the maximum
rotaton velocity of a galaxy (e.g. de Vaucouleurs et al. \cite{dV},
Tully et al. \cite{tt}). For the purposes of this discussion, we 
replace the axial ratio $a/b$ with 1/sec\,$i$ and take the extinction
parameter in the $B$ passband from Tully et al. (\cite{tt}):
\begin{center}
$C_B$\,=\,1.57\,+\,2.75\,[log(2\,V$_{\rm max}$)\,--\,2.5] 
\end{center}

Adopting V$_{\rm max}$=\,126  kms$^{-1}$ (see item 3.5) and $i=80^{\rm
o}  \pm  5^{\rm  o}$,  we  obtain   $A_B  =   0.^m99^{+0.39}_{-0.22}$.
Therefore, the total extinction-corrected  $B$-band absolute magnitude
of     ESO~603-G21      is      $M_B^0(0)=-19.3^{+0.4}_{-0.2}$      or
$L_B^0(0)=8.2\,10^9~L_{\odot,B}$.  On the other hand, the Tully-Fisher
relation predicts that the extinction-corrected luminosity of a galaxy
with V$_{\rm max}$=\,126 km/s is also  $M_B^0(0)=-19.3$  (Tully et al.
\cite{tt}).  Thus, one can conclude that  $A_B=1\fm0$  is a reasonable
estimate of the internal absorption in the galaxy and that ESO~603-G21
probably satisfies the Tully-Fisher relation for normal spirals.

The  relatively  high degree of symmetry  of this  object and its high
inclination  angle make it suitable  for a study of the dust lane.  In
order to estimate the extinction  law in the dust lane, we compare the
surface  brightness  of  regions  which are  equidistant  from the
nucleus on either  side along the minor  axis (see e.g.  Knapen et al.
\cite{k}).  Locating the exact center of the galaxy is very  important
for the asymmetry  study.  We adopt the center  position as determined
from the  $I$-band  image of  ESO~603-G21.  In Fig.~5 we  display  the
selective asymmetry (which is the difference between the unobscured NE
part of the minor axis profile and the SW part in one passband  versus
the same difference in another passband) at $r=3\arcsec-5\arcsec$ from
the  galaxy  center  (dust  obscuration  is  probably  present  in the
center).  The mean extinction relations for the dust lane are:

\begin{center}
$A_B$\,=\,(1.17\,$\pm$\,0.13)\,$A_V$, \\
$A_R$\,=\,(0.48\,$\pm$\,0.18)\,$A_V$, \\
$A_R$\,=\,(0.42\,$\pm$\,0.17)\,$A_B$. 
\end{center}

%---------------------------- Figure 4
\begin{figure}[!ht]
\psfig{file=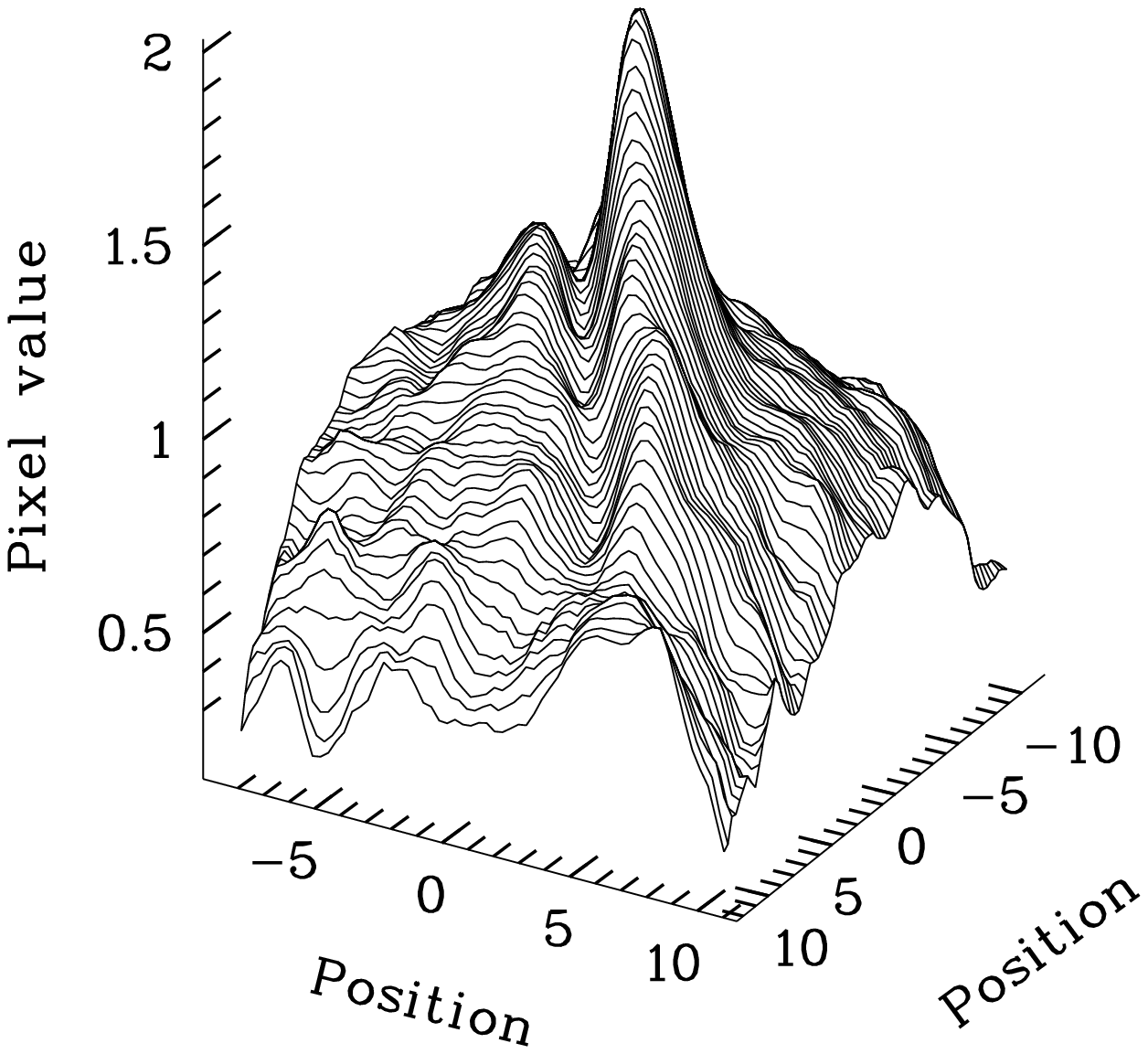,width=8.7cm,angle=-90,clip=}
\caption[4]{A 3-dimensional distribution of the $B-R$  color  index  
within the central ($\pm 10''$) part of ESO~603-G21. The orientation is
such that the line of sight coincides with the major axis of the 
galaxy. }
\end{figure}

The slopes of the selective asymmetry relations indicate that galactic
extinction law (given as solid straight  lines in Fig.~5) is valid, at
least as a first approximation, for the dust in ESO~603-G21.

\begin{figure*}
\psfig{file=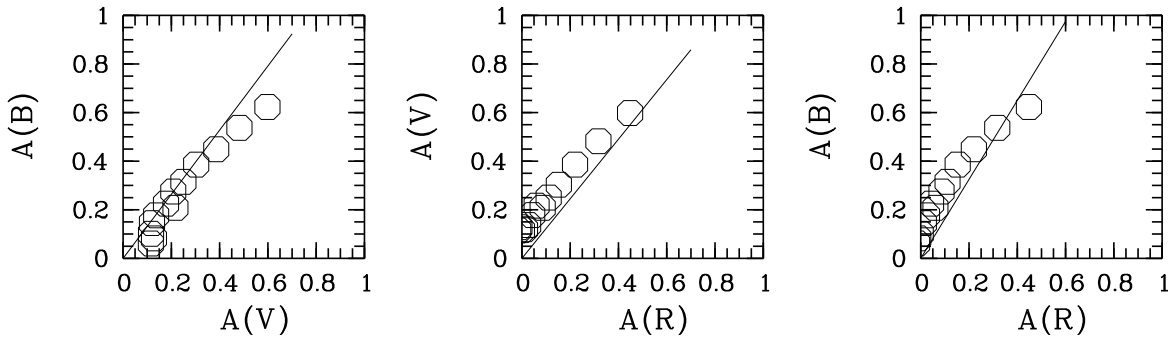,width=16.0cm,angle=-90,clip=}
\caption[5]{Selective asymmetry for ESO~603-G21 at 
$3\arcsec-5\arcsec$ from the nucleus (circles).
 The solid line represents the galactic 
standard extinction law.}
\end{figure*}

\subsection{Rotation curve}

The  emission-line  rotation curve for ESO~603-G21  along the apparent
major  axis  (P.A.=\,114$^{\rm  o}$) has  been  published  in PRC.  In
Fig.~6 we show the observed  rotation curve of the galaxy and our fit
by an exponential disk with $h$=1.22 kpc (Table 2) and intrinsic axial
ratio =\,0.1.  It is evident that the exponential  disk  approximation
gives a good description of the observed  rotation curve within 20$''$
from the nucleus.

\begin{figure}
\psfig{file=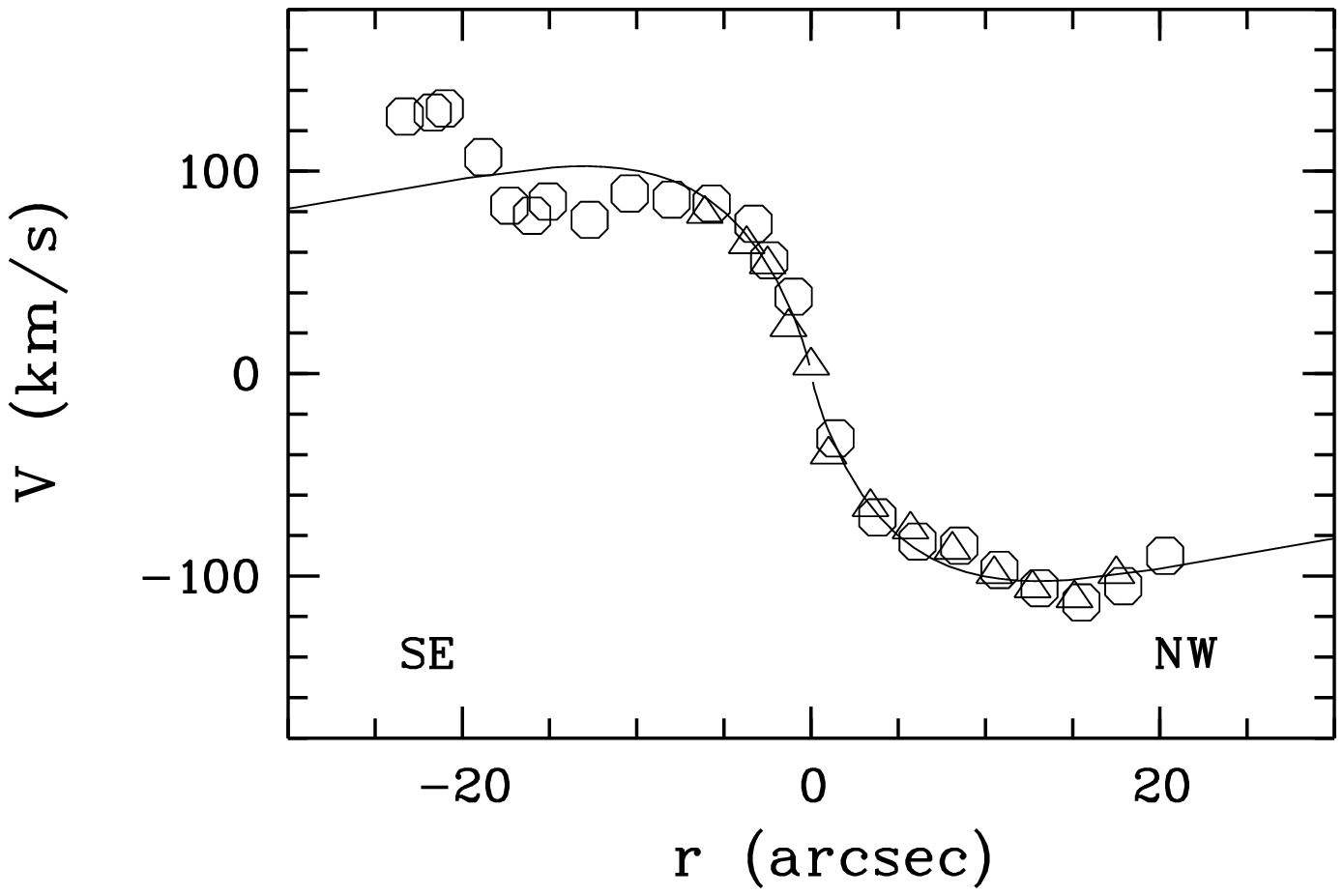,width=8.7cm,angle=-90,clip=}
\caption[6]{Emission-line  rotation  curve  along  the  major  axis of
ESO~603-G21   according  to  PRC.  The  circles  are  from   H$\alpha$
measurements,  the triangles  are from [NII]  measurements.  The solid
line shows the rotation curve of an exponential disk with scalelength 
of 1.22 kpc.}
\end{figure}

To obtain  the  global  maximum  rotation  velocity,  we have used the
following usual definition:

$$
{\rm V}_{\rm max} = W_{50}/2~(1+z)^{-1}({\rm sin}\,i)^{-1},
$$

\noindent where $W_{50}$ is the HI line width at 50\% of the peak, $z$
is the galaxy  redshift, and $i$ is the  inclination  angle (see Table
2).  We have obtained  V$_{\rm  max}$=126  kms$^{-1}$ and a total mass
within    the    optical    radius     (R$_{25}$\,=\,5.7    kpc)    of
2.1x10$^{10}$\,M$_{\odot}$  (assuming a spherical mass  distribution).
Thus,    the    mass-to-luminosity     ratio    is    M/$L_B^0(0)$=2.6
M$_{\odot}$/$L_{\odot,B}$, a value that is usual for disk galaxies.

\section{Discussion and conclusions}

The global  kinematical  structure of ESO~603-G21 -- stellar  rotation
along two orthogonal  position angles (Arnaboldi et al.  \cite{a1}) --
suggests  that the object is a polar-ring  galaxy.  The host galaxy is
probably  an  early-type  galaxy  with  an  exponential-like   surface
brightness distribution.  The central galaxy is surrounded by a warped
star-forming ring or disk.  In general,  ESO~603-G21  looks similar to
other classic PRG (e.g.  NGC~4650A).

There are, nonetheless,  several facts  complicating such an interpretation.
First, the central round component shows very low surface  brightness which
may indicate that the central  object is not an  early-type  galaxy  
like in most classic PRG (PRC). 
Second, in the  near-infrared  region  ($K$  passband)  most of the stellar
light  comes from a bright  nearly-exponential  disk.  Third,  the  central
round  object,  clearly  visible in the optical  images  (Fig.~2), is quite
faint in the $K$ passband  (Arnaboldi  et al.  \cite{a}).  The  totality of
the observed  characteristics  (optical and NIR colors, color gradients, HI
and H$_2$ content, FIR luminosity and star-formation  rate,  rotation-curve
shape -- Fig.~6 --, the agreement  with the  Tully-Fisher  relation,  etc.)
suggests that ESO~603-G21 could be an unusual late-type spiral galaxy with
a kinematically-decoupled  extended "bulge".  Therefore, it may be similar 
in some respects to
NGC~4672 and NGC~4698,  which are early-type  disk galaxies with  geometric
and kinematical  orthogonal  decoupling between the bulge and disk (Bertola
et  al.  \cite{b},   Sarzi  et  al.  \cite{sa}),  or to NGC~2748,
which is a late-type spiral galaxy with possible ongoing accretion of a dwarf
companion onto the central region of the galaxy (Hagen-Thorn et al. 1996).  
The bulge-like feature of ESO~603-G21 can be a "true" polar ring that 
is formed during almost  perpendicular  accretion  of an
early-type companion onto central region of a pre-existing disk galaxy.

Another interesting  interpretation  of  the  observed  ESO  603-G21
peculiarities is that the galaxy may be the result of a dissipative  merger
event  (this  scenario  was  proposed  recently  by Iodice  et al.  2001 to
explain  the   NGC~4650A   puzzles).  According  to  Bekki  (1997,   1998),
dissipative  galaxy  merging  with  a  near  polar  orbit  orientation  can
transform two late-type  spirals into one PRG.  In this  scenario, a spiral
galaxy  falling  from the polar  axis of the  target  galaxy  triggers  the
outwardly  propogating  density  wave in the  gaseous  disk  of the  victim
galaxy.  Then, gaseous dissipation and star formation  transform the victim
disk into polar ring or disk.  The central  object is the intruding  galaxy
that has been turned into an early-type-like galaxy during the merging.

\begin{figure}
\psfig{file=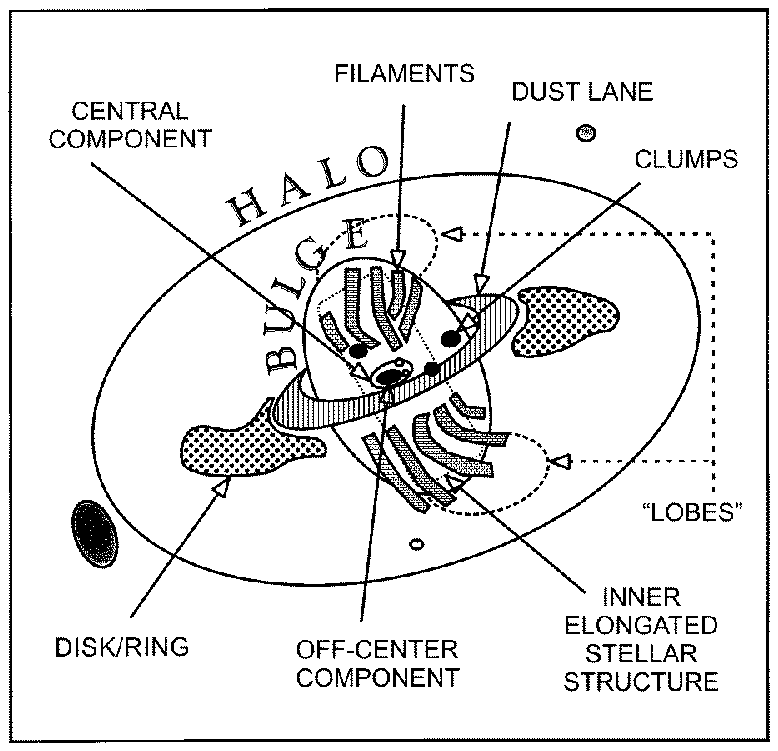,width=8.7cm,clip=}
\caption[7]{Sketch  of ESO 603-G21 as seen in careful visual inspection
of the images in the various  filters. North
is up and East is on the left.}
\label{draw}
\end{figure}

Figure ~\ref{draw}  depicts  the
various  internal  substructures  of ESO 603-G21 as  revealed  in this
work (see item 3.2). Such complex, non-settled, fine structure of the
galaxy  supports our supposition about relatively late formation
of the ''bulge'' due to external accretion or a merger.

Interestingly  enough, the companion of ESO 603-G21 is \mbox{ESO  603-G20},
an edge-on  disk-galaxy  without any ``explicit"  evidence of  interaction.
The relative  velocity  between both objects is 65 kms$^{-1}$  (see NED and
references  therein);  this  suggests  that both objects may form a
bound system!  There is, nonetheless, a third
faint  object  between  them,  which  seems to bear a very faint and narrow
bridge to  ESO~603-G21.  The  coordinates  of the centroid  (J2000) of this
object are $\alpha$ = 22$^{\rm h}$ 51$^{\rm m}$ 10.4$^{\rm s}$ and $\delta$
=  -20$^{\rm  o}$  14$'$  59.5$\arcsec$  within  1$\arcsec$  of  error,  as
calculated  from  the  Digitized  Sky  Survey  (DSS)  image  (see  Fig.~8).
Therefore,   we  have   denoted   this   object   
Anon~J225110.4-201459.5.  This
is probably a low-surface brightness galaxy.  On the basis of the DSS image
we  have  found  that  the  $B$-band  total  magnitude  of  the  galaxy  is
$B_T=18.0\pm0.5$.

\begin{figure}
\psfig{file=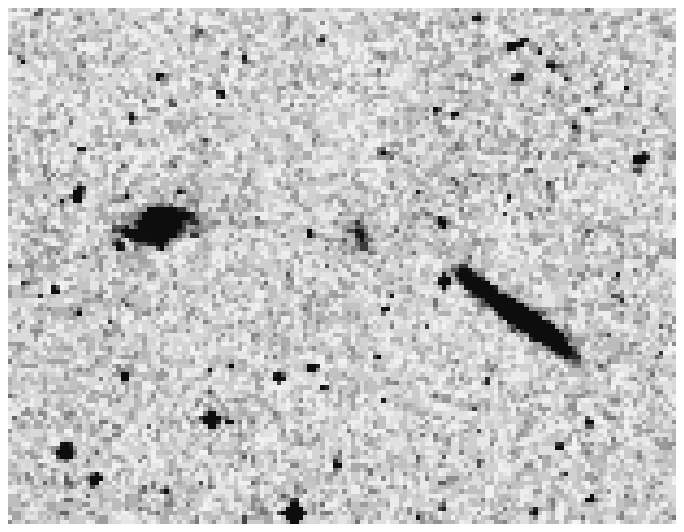,width=8.7cm,clip=}

\caption[8]{In  spite of the DSS  resolution,  a  high-pass  filtering  has
slightly enhanced the probable bridge between ESO 603-G21 (on the left) and
Anon~J225110.4-201459.5  at the center (see text).  ESO 603-G20 is on
the right. North is to the top, east on the left.}

\label{dss}
\end{figure}

It is essential to note that the disks of  \mbox{ESO  603-G21}  (Fig.1) and
\mbox{ESO 603-G20}  (Fig.~\ref{dss}) are strongly warped.  This feature, as
well as the probable bridge, may be an indication of ongoing interaction in
the \mbox{ESO  603-G21}--\mbox{ESO   603-G20}--Anon~J225110.4-201459.5   triple
system (e.g.  Reshetnikov \& Combes 1999).  So \mbox{ESO 603-G21} is not an
isolated  object, but a member of a group of galaxies  (like, for instance,
NGC~4650A).  Such dense spatial  environment  supports the 
idea that  \mbox{ESO  603-G21}  may  represent  the  result of  a
merging event.  To test this scenario, detailed  numerical  simulations are
needed.

\acknowledgements{We  would like to thank the  referee,  Dr.  G.  Galletta,
for useful comments and  suggestions.  V.R.  acknowledges  partial  support
from the  ``Integration"  programme  (A0145) and the DAAD (Germany).  M.F-A
and M.  de O-A thank the partial support of the  Funda\c{c}\~{a}o de Amparo
\`{a} Pesquisa do Estado de Minas Gerais  (FAPEMIG,  grant CEX 1864/95) and
the   Minist\'{e}rio  da  Ci\^{e}ncia  e  Tecnologia  (MCT,  Brazil).  This
research has made use of the NASA/IPAC  Extragalactic  Database (NED) which
is operated by the Jet Propulsion  Laboratory, Caltech, under contract with
the  National  Aeronautics  and Space  Administration.  The  Digitized  Sky
Survey was produced at the Space Telescope Science Institute (ST ScI) under
U.S.  Government  grant NAG  W-2166.  Also, this  publication  makes use of
data products from the Two Micron All Sky Survey, which is a joint  project
of the University of Massachusetts and the Infrared Processing and Analysis
Center/California   Institute  of   Technology,   funded  by  the  National
Aeronautics  and  Space   Administration  and  the  National  Science  
Foundation.  }

\end{document}